\documentclass[conference]{IEEEtran}
\usepackage{flushend}
\usepackage{epsfig}
\usepackage{graphicx}
\usepackage{amsmath}
\usepackage{amssymb}
\usepackage{verbatim}
\usepackage{mathrsfs}
\usepackage{lipsum}
\usepackage{graphicx}
\usepackage{subcaption}
\usepackage{amsmath,amssymb}
\usepackage{amssymb}%
\usepackage{mathrsfs}%
\usepackage{blindtext}
\usepackage{xcolor}
\usepackage{colortbl}
\usepackage{cite}
\IEEEoverridecommandlockouts
\def\BibTeX{{\rm B\kern-.05em{\sc i\kern-.025em b}\kern-.08em
    T\kern-.1667em\lower.7ex\hbox{E}\kern-.125emX}}
\begin{document}
\title{Channel Agnostic End-to-End Learning based Communication Systems with Conditional GAN}
\author{\IEEEauthorblockN{Hao Ye, Geoffrey Ye Li, and Biing-Hwang Fred Juang  
}\\

\IEEEauthorblockA{School of Electrical and Computer Engineering\\
Georgia Institute of Technology\\
Email: yehao@gatech.edu; liye@ece.gatech.edu; juang@ece.gatech.edu}

\and
\IEEEauthorblockN{Kathiravetpillai Sivanesan}\\
\IEEEauthorblockA{Intel Corporation \\
kathiravetpillai.sivanesan@intel.com}

\thanks{This work was supported by the National Science Foundation under Grants
1443894 and 1731017.}
}
\maketitle

\begin{abstract}
In this article, we use deep neural networks (DNNs) to develop a wireless end-to-end communication system, in which DNNs are employed for all signal-related functionalities, such as encoding, decoding, modulation, and equalization. However, accurate instantaneous channel transfer function, \emph{i.e.}, the channel state information (CSI), is necessary to compute the gradient of the DNN representing. In many communication systems, the channel transfer function is hard to obtain in advance and varies with time and location.
In this article, this constraint is released by developing a channel agnostic end-to-end system that does not rely on any prior information about the channel. We use a conditional generative adversarial net (GAN) to represent the channel effects, where the encoded signal of the transmitter will serve as the conditioning information. 
In addition, in order to deal with the time-varying channel, the received signal corresponding to the pilot data can also be added as a part of the conditioning information.
From the simulation results, the proposed method is effective on additive white Gaussian noise (AWGN) and Rayleigh fading channels, which opens a new door for building data-driven communication systems.

\end{abstract}

\section{Introduction}

\begin{figure}[!t]
\centering
\includegraphics[width=0.9\linewidth]{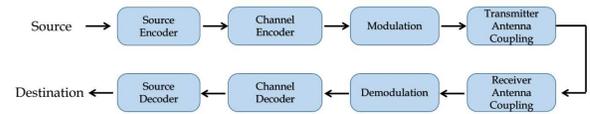}
\caption{A typical block diagram of digital communication systems.} \label{fig:commblocks}
\end{figure}

Many of our daily communication services rely on the digital communication technology, a block diagram of which is depicted in Fig. \ref{fig:commblocks}. 
While the technologies in this system are quite mature, individual blocks therein are separately designed, often with different assumptions and objectives, making it difficult if not impossible to ascertain global optimality of the system. In addition, the channel propagation is expressed in an assumed mathematical model embedded in the design. The assumed model may not be correctly reflected in the actual transmission scenario, thereby compromising the system performance.

Recently, deep learning has been utilized for improving the performance of  the traditional block-structure communication system, including the multiple-input and multiple-output (MIMO) detection \cite{MIMO_Det}, channel decoding \cite{Decoding_DNN}, and channel estimation \cite{Channel_DNN}. In addition, deep learning based methods also show potential improvement by jointly optimizing the processing blocks, including joint channel estimation and detection \cite{Hao}, joint channel encoding and source encoding \cite{Source_Channel_Coding}.

Besides enhancing the traditional communication blocks, deep learning provides a new paradigm of the communication system. As a pure data-driven method, the features and the parameters of a deep learning model can be learned directly from the data without handcraft or ad-hoc designs by optimizing an end-to-end loss function. Inspired by this methodology, end-to-end learning based communication systems have been investigated in several prior works \cite{physical_layer,Air, ofdm_end_to_end,RL_E2E} where both the transmitter and the receiver are represented as deep neural networks (DNNs) and can be interpreted as an auto-encoder and an auto-decoder, respectively.

\begin{figure}[!t]
\centering
\includegraphics[width=0.8\linewidth]{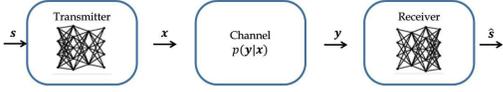}
\caption{The structure of the end-to-end learning based communication system.} \label{fig:e2estructure}
\end{figure}

The structure of the end-to-end learning based communication system is depicted  in Fig. \ref{fig:e2estructure}. As shown in the figure, the transmitter learns to encode the transmitted symbols into encoded data, $\mathbf{x}$, which is then sent to the channel, while the receiver learns to recover the transmitted symbols based on the received signal, $\mathbf{y}$, from the channel.

The weights of the model are trained in a supervised learning manner to optimize the end-to-end recovery accuracy. This idea is first proposed in \cite{physical_layer} and has been shown to have a similar performance as the traditional approaches with block structures under the additive white Gaussian noise (AWGN) channel. In \cite{Air}, the end-to-end method has been extended to handle the various hardware imperfection. In \cite{ofdm_end_to_end}, an end-to-end learning method is adopted within the orthogonal frequency-division multiplexing (OFDM) system. 


Despite these applications, a major limitation of the above end-to-end design paradigm is the instantaneous channel state information (CSI), $\mathbf{h}$, or the channel transfer function, $\mathbf{y} = f_\mathbf{h}(\mathbf{x})$, must be known when optimizing the transmitter.
As is well known, the weights of the DNN are usually updated using stochastic gradient descent (SDG) with the computed error gradients propagated from the output layer back to the input layer. However, when the channel in unknown, the back-propagation of the gradients is blocked by the unknown channel, preventing the overall learning of the end-to-end network. 
The channel transfer function may be assumed, but any such assumption would bias the learned weights, repeating the pitfalls caused by the likely discrepancy between the assumed model and the actual channel. 
In addition, in real communication systems, an accurate instantaneous CSI is hard to obtain in advance because the end-to-end channel often includes several types of random effects, such as channel noise and varying, which may be unknown or can not be expressed analytically. 
As a result, it is desirable to develop a channel agnostic end-to-end learning based communication system, where different types of channel effects can be automatically learned without knowing the specific channel transfer function. 

In order to achieve this goal, two major challenges should be addressed. First, since the back-propagation is blocked by of the unknown channel, one needs to attempt to train the transmitter with surrogate gradients or without gradients. Second, in many communication systems, the channels may vary with time and location but the instantaneous CSI is vital for coherent detection. Recently, reinforcement learning has been employed in \cite{RL_E2E} to optimize an end-to-end transmitter without knowing the channel transfer function, where the channel and the receiver are considered as the environment when training the transmitter. But some prior information of the channel is still needed to achieve a competitive performance.

In this article, we propose a channel agnostic end-to-end learning based communication system where the distribution of channel output can be learned through a conditional generative adversarial net (GAN) \cite{conditional}. The conditioning information is the encoded signals from the transmitter along with the received pilot information used for estimating the channel.  By iteratively training the conditional GAN, the transmitter, and the receiver, the end-to-end loss can be optimized in a supervised way. A concurrent work \cite{void} has modeled the channel effects with a GAN and is similar to our work in that sense. But we build an end-to-end model and our proposed approach can be applied to more realistic time-varying channels, which is significantly different from the existing work.

Our main contributions are three-folded. First, we use the conditional GAN to model the channel conditional distribution, $p(\mathbf{y}|\mathbf{x})$, so that the channel effects can be learned based on the data instead of expert knowledge about the channel. Second, by adding the pilot information as a part of the conditioning information for the time-varying channels, the conditional GAN can generate more specific samples for the current channel. 
Third, an end-to-end learning based communication system is developed, where the gradients of the end-to-end loss can be propagated to the transmitter through the conditional GAN.

The rest of the paper is organized as follows. In Section \ref{sec:GAN}, the conditional GAN based channel modeling approach is introduced. In Section \ref{sec:End2End}, the training for the end-to-end system is presented in detail. In Section \ref{sec:Exp}, the simulation results are presented and the conclusions are drawn in Section \ref{sec:Conclusion}.

\section{Modeling Channel with Conditional GAN} \label{sec:GAN}


The end-to-end communication system learns the implements of the transmitter and the receiver using DNNs. However, the back-propagation, which is used to train the weights of DNNs, is block by the channel, preventing the overall learning of the end-to-end network. 
To address the issue, we use a conditional GAN to learn the channel effects and the learned model can act as a bridge for the gradients to pass through. 
In this section, the conditional GAN is introduced and the how to use the conditional GAN to model the channel effects is presented.

\subsection{Conditional GAN}

\begin{figure}[!t]
\centering
\includegraphics[width=0.9\linewidth]{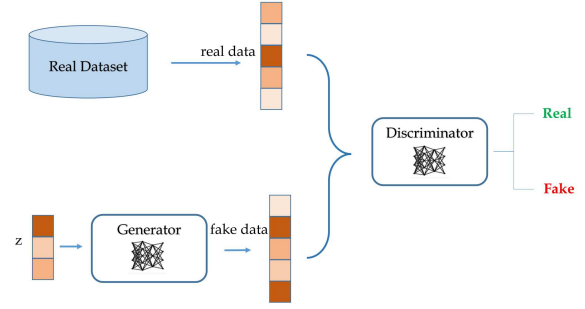}
\caption{Structure of GAN.} \label{fig:GAN}
\end{figure}

GAN is a new class of generative methods for distribution learning, where the objective is to learn a model that can produce samples close to some target distribution, $p_{data}$. In our system, a GAN is applied to model the distribution of the channel output and the learned model is then used as a surrogate of the real channel when training the transmitter so that the gradients can pass through to the transmitter.

The structure of the GAN is shown in Fig. \ref{fig:GAN},  where a min-max two players game is introduced between a generator, $G$, and a discriminator, $D$. The discriminator, $D$, learns to distinguish between the data generated by the generator and the data from the real dataset while the generator, $G$, learns to generate samples to fool the discriminative networks into making mistakes.

During the training, the generator maps an input noise, $\mathbf{z}$, with prior distribution, $p_z(\mathbf{z})$, to a sample. Then the samples from the real data and the samples generated from the generator, $G$, are collected to train the discriminator, $D$, to maximize the ability to distinguish between the two categories. If the discriminator, $D$, is successful at classifying the samples of the two sources, then its success can be used to generate a feed back to the generator, $G$, so that the generator, $G$, will learn to produce samples more similar to the real samples. The training procedure will end when reaching the equilibrium, where the discriminator, $D$, can do no better than random guessing to distinguish the real samples and the generated fake samples.

Both the generator, $G$, and the discriminator, $D$, are represented by a DNN, with parameters $\mathcal{G}$ and $\mathcal{D}$, respectively, and the objective for optimization is 

\begin{equation}
\begin{split}
\min_{\mathcal{G}}\max_{\mathcal{D}} V(D, G) =& E_{\mathbf{x}\sim p_{data}(\mathbf{x})} [\log(D_{\mathcal{D}}(\mathbf{x}))] \\
&+ E_{\mathbf{z}\sim p_z(\mathbf{z})}[\log(1 - D_{\mathcal{D}}(G_{\mathcal{G}}(\mathbf{z})].
\end{split}
\end{equation}

The object of the discriminator, $D$, is to give a high value when the input belongs to the real dataset and a low value when the input is generated by the generator, $G$, while the object of generator, $G$, is to maximize the output of the discriminator, $D$, given the generated samples, $G(\mathbf{z})$.  

\begin{figure}[!t]
\centering
\includegraphics[width=0.9\linewidth]{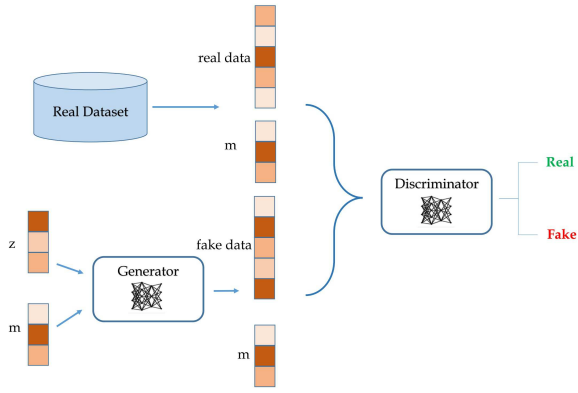}
\caption{Structure of conditional GAN.} \label{fig:CGAN}
\end{figure}

The GAN can be extended to a conditional model if both the generator, $G$, and the discriminator, $D$, are conditioned on some extra information, $\mathbf{m}$.  The structure of the conditional GAN is shown in Fig. \ref{fig:CGAN}. We only need to feed the conditioning information, $\mathbf{m}$, into both the generator, $G$, and discriminator, $D$, as the additional input. Therefore, the output of the generator, $G$, will be $G(\mathbf{x}|\mathbf{m})$ and the output of discriminator, $D$, will be $D(\mathbf{x}|\mathbf{m})$.  The min-max optimization objective becomes
\begin{equation}
\begin{split}
\min_{\mathcal{G}}\max_{\mathcal{D}} V(D, G) &= E_{\mathbf{x}\sim p_{data}(\mathbf{x})} [\log(D_{\mathcal{D}}(\mathbf{x}|\mathbf{m}))] \\
&+ E_{\mathbf{z}\sim p_z(\mathbf{z})}[\log(1 - D_{\mathcal{D}}(G_{\mathcal{G}}(\mathbf{z}|\mathbf{m})].
\end{split}\label{fig:E2E}
\end{equation} 
The conditional GAN is employed in our end-to-end system to model the channel output distribution with given conditioning information on the encoded signal and the received pilot data.

\begin{figure}[!t]
\centering
\includegraphics[width=0.8\linewidth]{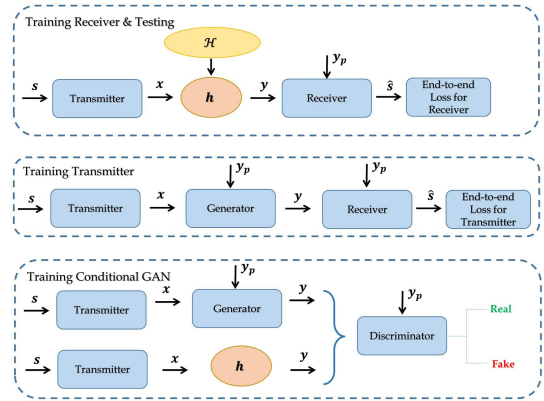}
\caption{Training and testing of the end-to-end system.} \label{fig:E2E}
\end{figure}

\subsection{Modeling Channels}

GAN is a powerful tool in learning distribution and the channel output, $\mathbf{y}$, given input, $\mathbf{x}$, is determined by the conditional distribution, $p(\mathbf{y}|\mathbf{x})$. Therefore, a conditional GAN can be employed for learning the output distribution of a channel by taking the $\mathbf{x}$ as the condition information. The generator will try to produce the samples similar to the output of the real channel while the discriminator will try to distinguish data coming from the real channel and the data coming from the generator.

The instantaneous CSI, $\mathbf{h}$, can be regarded as a sample from a large channel set, $\mathcal{H}$ and is also vital coherent detection of the data at the receiver. In order to obtain the CSI, a common practice is to send some pilot information to the receiver so that the channel information is inferred based on the received pilot information, $\mathbf{y_p}$.  In our proposed method, the received pilot information, $\mathbf{y_p}$, can be added as a part of the conditioning information so that the output samples follow the distribution of $\mathbf{y}$ give the $\mathbf{x}$ and the received pilot data, $\mathbf{y_p}$.

\section{End-to-End Communication System} \label{sec:End2End}

With the conditional GAN, the gradients can be back-propagated to the transmitter. Following the previous works\cite{physical_layer}, the transmit symbol drawn from a finite discrete set of size $M$ is converted into a one-hot vector, $\mathbf{s}$, of length $M$ and the end-to-end transmission is treated as a $M$-class classification problem. The output of the receiver, $\mathbf{\hat{s}}$, is a probability vector over the $M$ possible classes. The cross-entropy loss is computed at the receiver, which is defined as
\begin{equation}
L = \sum_{n=1}^{M} -s_n\log (\hat{s}_n),
\end{equation}
where $s_n$ and $\hat{s}_n$ represent the $n$th elements of $\mathbf{s}$ and $\mathbf{\hat{s}}$, respectively.

The training and testing of the proposed end-to-end communication system are shown in Fig. \ref{fig:E2E}. During the training, the data are obtained, where the transmitted symbols are randomly generated and the instantaneous CSI is sampled randomly from the channel set. Based on the training data, the transmitter, the receiver, and the channel generator in the conditional GAN can be trained iteratively and when training one component, the parameters of the others remain fixed. When training the receiver and the transmitter, the object is to minimize the end-to-end loss. The object is to minimize the min-max optimization objective when training the conditional GAN for generating the channel.
In the testing stage, the end-to-end reconstruction performance is evaluated on the learned transmitter and the receiver with real channels.

\subsection{Training Receiver}
The receiver can be trained easily since the loss function is computed at the receiver, thus the gradients of the loss can be easily obtained. The input of the DNN will be the received signal, $\mathbf{y}$, and the receive pilot data, $\mathbf{y_p}$.  For the time-varying channels, by directly put the received signal, $\mathbf{y}$, and the receive pilot data, $\mathbf{y_p}$, together as the input, the receiver can automatically infer the channel condition and perform the channel estimation and detection simultaneously without explicitly estimating the channel. 

\subsection{Training Transmitter}
With the channel generator being a surrogate channel, the training of the transmitter will be similar to the training of the receiver. The end-to-end cross-entropy loss is computed at the receiver, and the gradients are propagated back the transmitter through the conditional GAN. The weights of the transmitter will be updated based on SGD while the weights of the conditional GAN and the receiver remain fixed.

\subsection{Training Channel Generator}
The channel generator is trained with the discriminator together. With the learned transmitter, the real data can be obtained with the encoded signal from the transmitter going through the real channel while the fake data is obtained from the encoded data going through the channel generator. The objective function to optimize is shown in Equation (2).

\section{Experiments} \label{sec:Exp}

In this section, simulation results on the AWGN channel and the Rayleigh fading channel are provided. We compare our channel agnostic learning based approach with the traditional methods, which are designed based on the channel transfer functions.

The structures and parameters of each model are listed in Table \ref{tab:Model}. The weights are updated by Adam \cite{ADMM} and the batch size for training is 320.

\begin{table}
\centering
\caption{Model Parameters}
\begin{tabular}{|c|c|}
\hline
\rowcolor{gray}
Parameters & Values \\ \hline
Transmitter hidden layers & {32, 32} \\ \hline
Learning rate & 0.001 \\ \hline
Receiver hidden layers & {32, 32} \\ \hline
Learning rate & 0.001 \\ \hline
Generator hidden layers & {128, 128, 128} \\ \hline
Discriminator hidden layers & {32, 32, 32} \\ \hline
Learning rate & 0.0001 \\ \hline
\end{tabular}\label{tab:Model}
\end{table}

\subsection{AWGN Channel}
The proposed method is first applied in the AWGN channel, where the output of the channel, $\mathbf{y}$, is the summation of the input signal, $\mathbf{x}$, and Gaussian noise, $\mathbf{w}$, that is, $\mathbf{y} = \mathbf{x} + \mathbf{w}$. In this case, there is no need for channel estimation. Thus the conditioning information is only the encoded signal from the transmitter.

\begin{figure}[!t]
\centering
\includegraphics[width=0.8\linewidth]{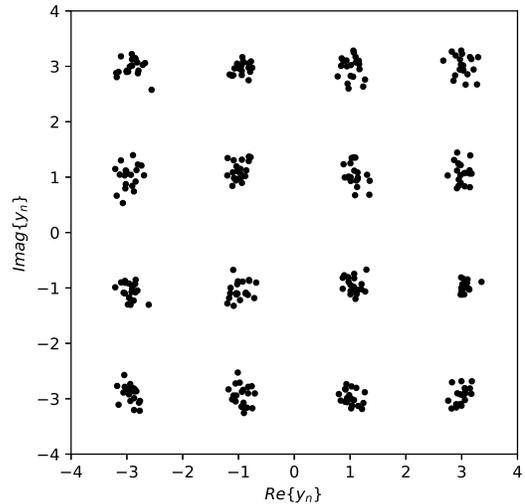}
\caption{Signal constellations at the output of an AWGN channel represented by a conditional GAN.} \label{fig:AWGN_mod}
\end{figure}

We first test the ability of the conditional GAN in learning the channel output distribution. Fig. \ref{fig:AWGN_mod} shows the output of the channel generator with the standard 16 QAM modulation as the conditioning information.  From the figure, the synthetic samples produced by the generator are very similar to the output from the AWGN channel.

\begin{figure}[!t]
\centering
\includegraphics[width=0.9\linewidth]{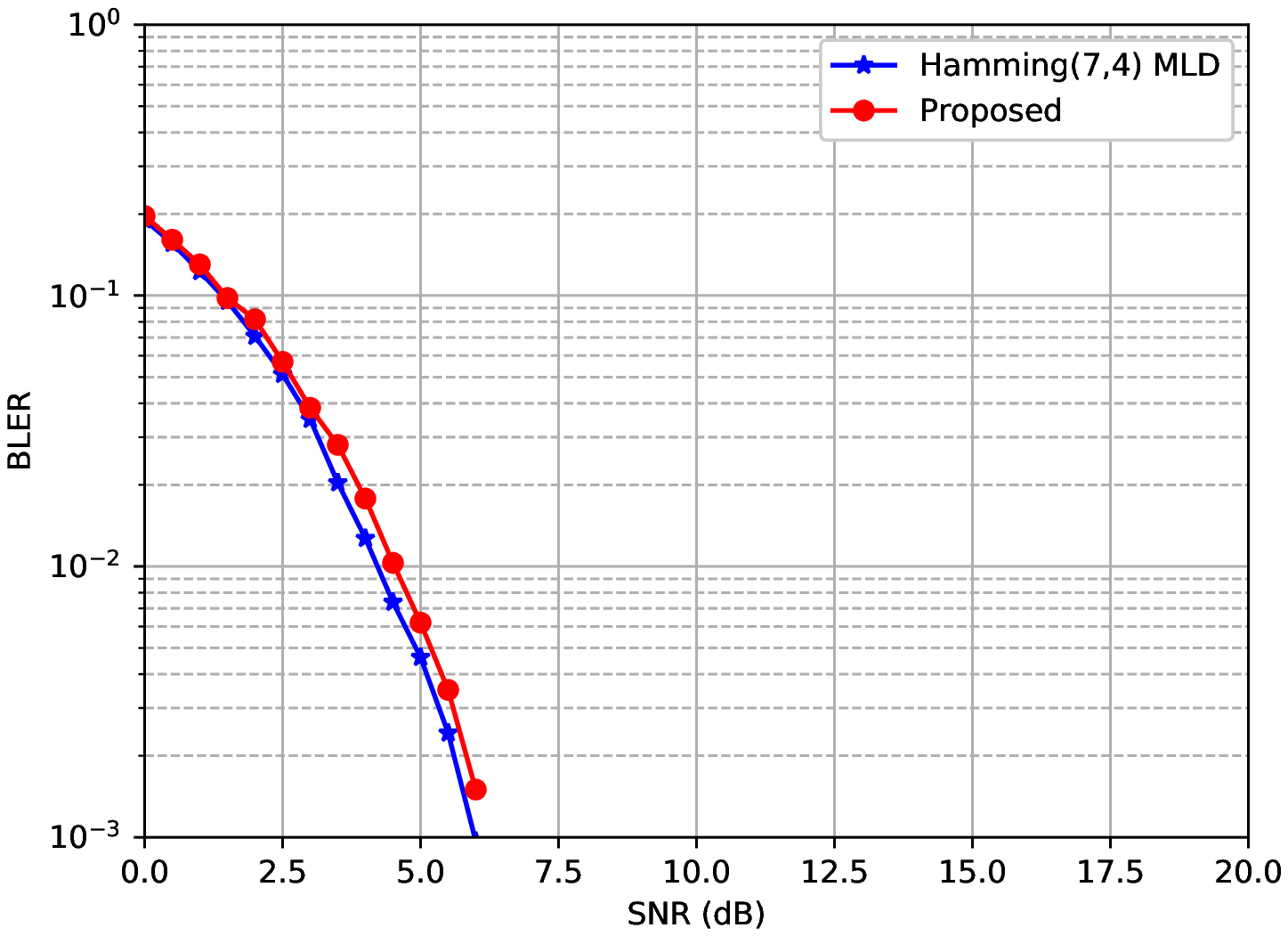}
\caption{BLER vs SNR.} \label{fig:AWGN_ber}
\end{figure}

The end-to-end recovering performance on the AWGN channel is shown in Fig. \ref{fig:AWGN_ber}. At each time, four information bits are transmitted and  the length of the transmitter output is set to be seven. From the figure, the block-error rate (BLER) of learning based approach is similar to Hamming (7,4) code with maximum-likelihood decoding (MLD).

\subsection{Rayleigh Fading Channel}

In the case of the Rayleigh fading channels, the channel output is determined by $y_n = h_n \cdot x_n + w_n$, where $h_n \sim \mathcal{CN}(0,1)$. Since the channel is time-varying, additional conditional information is added to the channel generator and the receiver. We can use real $\mathbf{h}$ for coherent detection task or received pilot data, $\mathbf{y_p}$, for joint channel estimation and detection (the pilot is assumed to be 1). 

\begin{figure}[ht] 
  \begin{subfigure}[b]{0.5\linewidth}
    \centering
    \includegraphics[width=1\linewidth]{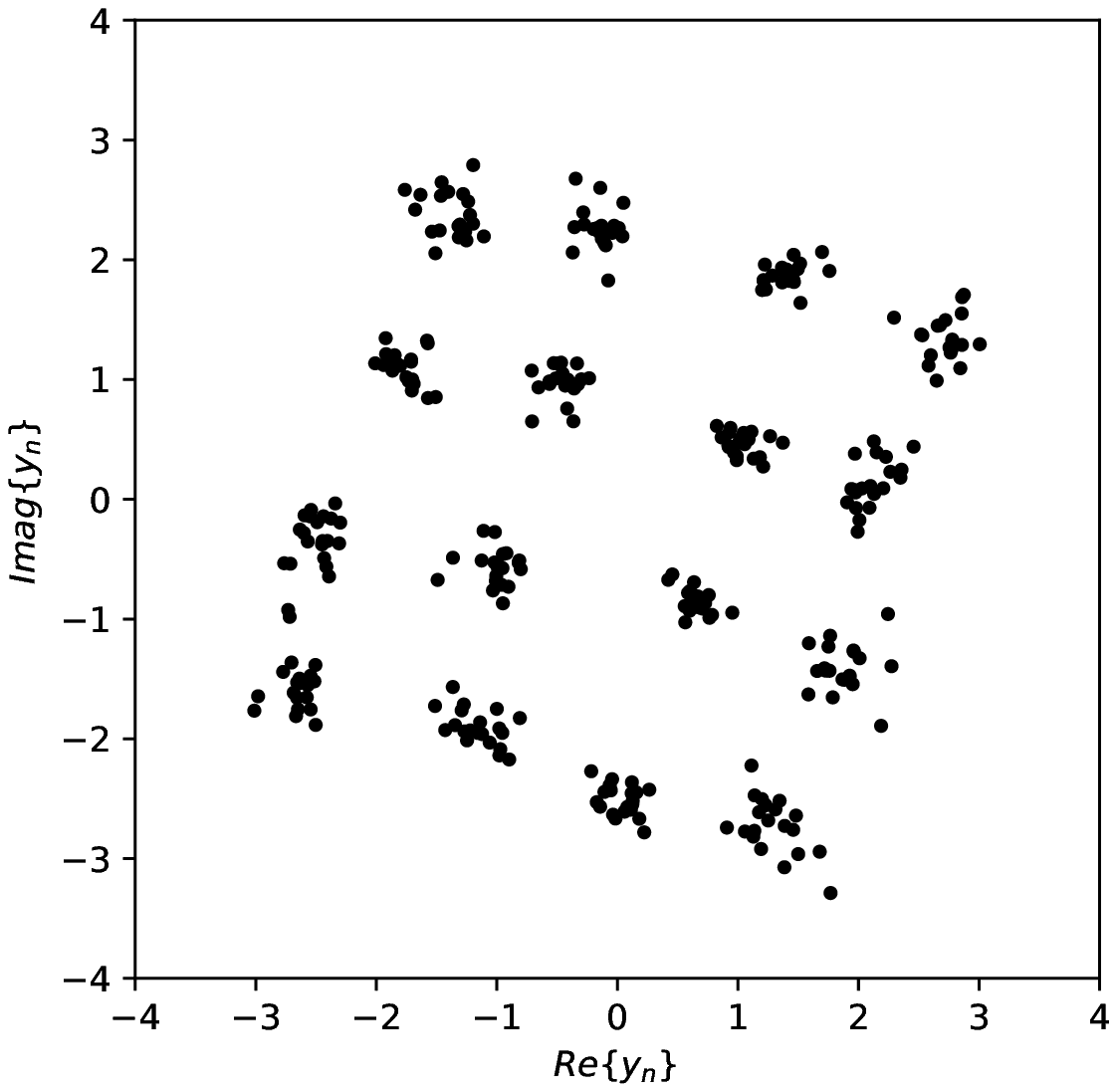} 
  \end{subfigure}
  \begin{subfigure}[b]{0.5\linewidth}
    \centering
    \includegraphics[width=1\linewidth]{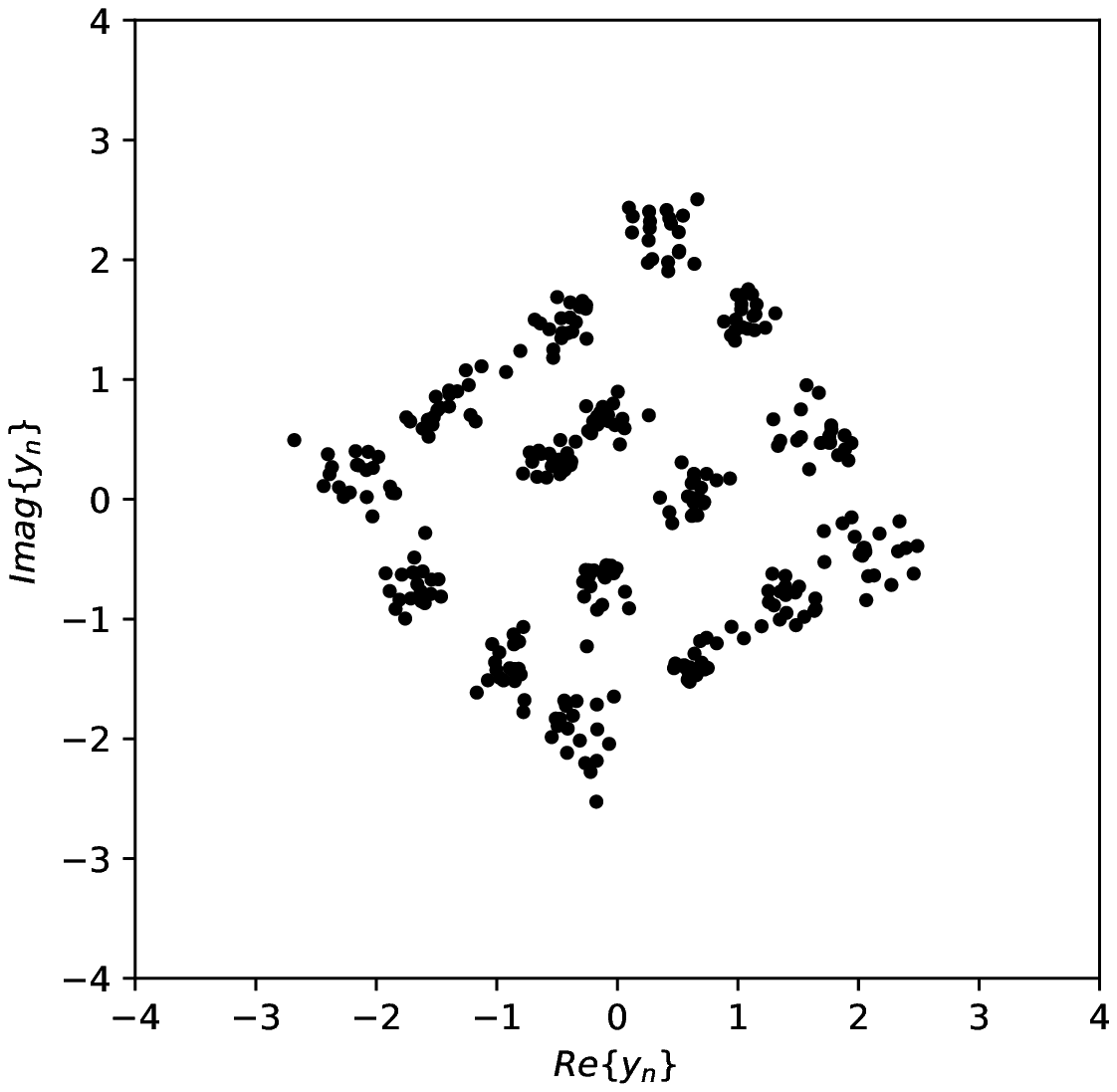} 
  \end{subfigure} 
  \caption{Signal constellations at the output of a Rayleigh channel represented by a conditional GAN}
  \label{fig:Ray_mod} 
\end{figure}

We first test the effectiveness of conditional GAN in learning the distribution of the channel with standard 16 QAM as the encoded symbols. Fig. \ref{fig:Ray_mod} shows generated samples with different channel values added to the conditioning information. From the figure, the conditional GAN is able to produce the samples with various means according to conditioning information.

\begin{figure}[!t]
\centering
\includegraphics[width=0.9\linewidth]{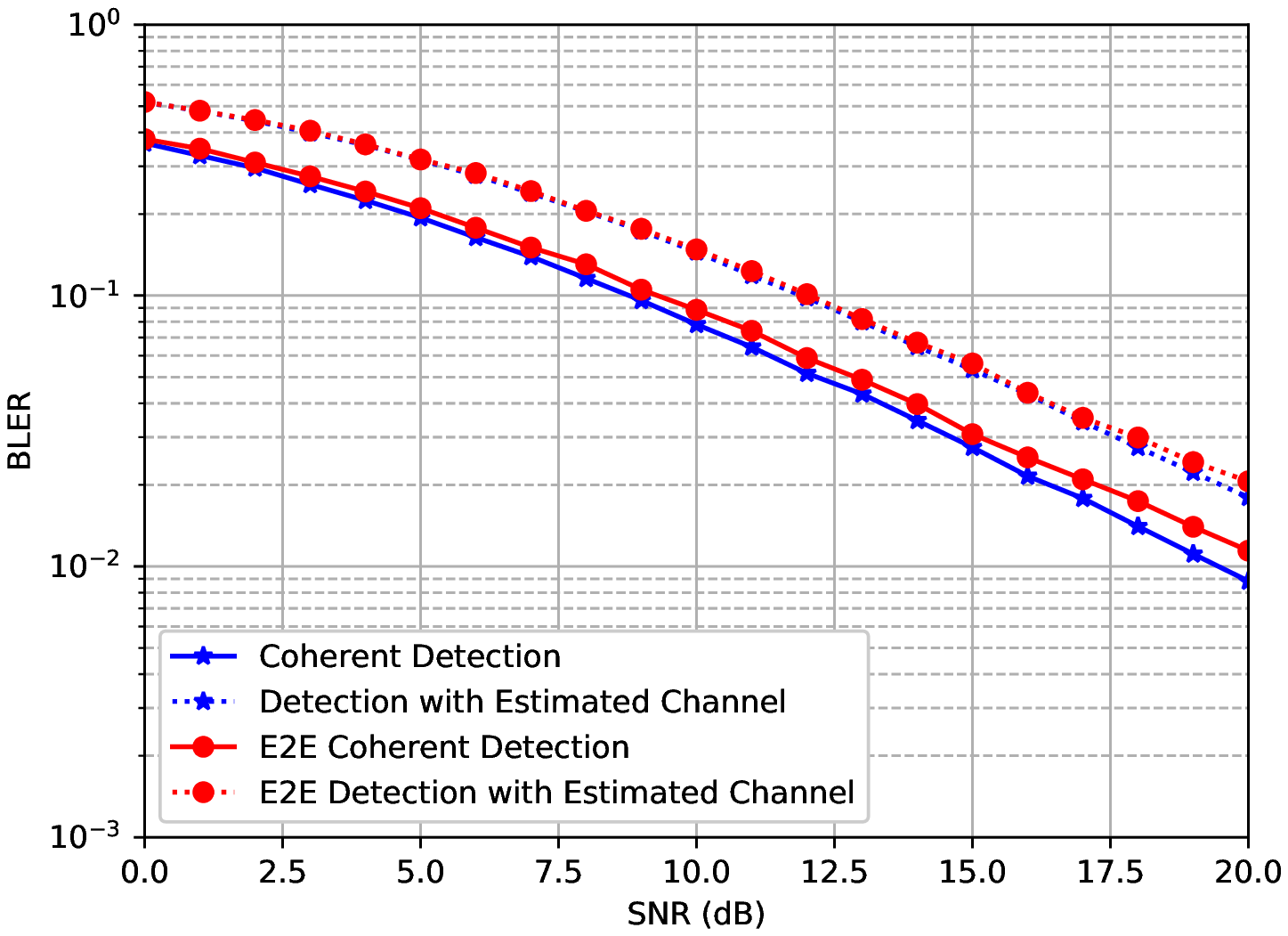}
\caption{BLER vs SNR.} \label{fig:Ray_ber}
\end{figure}

The end-to-end reconstruction results are shown in Fig. \ref{fig:Ray_ber}, where the learning based end-to-end approach shows a similar performance to the traditional methods in both tasks.

\section{Conclusions} \label{sec:Conclusion}
DNN has been used to develop end-to-end communication systems, where both the transmitter and the receiver are represented by DNNs. The accurate instantaneous channel transfer function is necessary to compute the gradients for optimizing the transmitter. However, in many communication systems, the channel transfer function is hard to obtain and varies with time and location.
In this article, we show that the conditional distribution of the channel can be modeled by the conditional GAN, which enables the end-to-end learning of a communication system without prior information of the channel. In addition, by adding the pilot information into the condition information, the conditional GAN can generate data corresponding to the specific instantaneous channel.

The end-to-end pipeline consists of DNNs for the transmitter, the channel generator, and the receiver. By iteratively training these networks, the end-to-end loss can be optimized in a supervised way. The simulation results confirm the effectiveness of the proposed method, by showing similar performance with traditional approaches based on expert knowledge and channel models.

Our method provides a new way for the end-to-end learning system. Although the simulation is only based on the AWGN and the Rayleigh fading channels, it can be easily extended to other channels, therefore opens a new door for building the pure data-driven communication systems.

\bibliographystyle{IEEEtran}
\bibliography{e2e}
\bibliography{strings,refs}

\end{document}